\documentclass[prb,twocolumn, superscriptaddress, nolongbibliography]{revtex4-2}

\usepackage[dvipdfmx]{graphicx}
\usepackage{dcolumn}
\usepackage{bm}
\usepackage{color}

\begin{document}

\title{Photo-induced nonlinear band shift and valence transition in SmS}
\author{Yitong Chen}
\affiliation{Department of Physics, Graduate School of Science, Osaka University, Toyonaka, Osaka 560-0043, Japan}
\author{Takuto Nakamura}
\email{nakamura.takuto.fbs@osaka-u.ac.jp}
\affiliation{Graduate School of Frontier Biosciences, Osaka University, Suita, Osaka 565-0871, Japan}
\affiliation{Department of Physics, Graduate School of Science, Osaka University, Toyonaka, Osaka 560-0043, Japan}
\author{Hiroshi Watanabe}
\email{hwata.fbs@osaka-u.ac.jp}
\affiliation{Graduate School of Frontier Biosciences, Osaka University, Suita, Osaka 565-0871, Japan}
\affiliation{Department of Physics, Graduate School of Science, Osaka University, Toyonaka, Osaka 560-0043, Japan}
\author{Takeshi Suzuki}
\affiliation{Institute for Solid State Physics, The University of Tokyo, Kashiwa, Chiba 277-8581, Japan}
\author{Qianhui Ren}
\affiliation{Institute for Solid State Physics, The University of Tokyo, Kashiwa, Chiba 277-8581, Japan}
\author{Kecheng Liu}
\affiliation{Institute for Solid State Physics, The University of Tokyo, Kashiwa, Chiba 277-8581, Japan}
\author{Yigui Zhong}
\affiliation{Institute for Solid State Physics, The University of Tokyo, Kashiwa, Chiba 277-8581, Japan}
\author{Teruto Kanai}
\affiliation{Institute for Solid State Physics, The University of Tokyo, Kashiwa, Chiba 277-8581, Japan}
\author{Jiro Itatani}
\affiliation{Institute for Solid State Physics, The University of Tokyo, Kashiwa, Chiba 277-8581, Japan}
\author{Shik Shin}
\affiliation{Office of University Professor, The University of Tokyo, Kashiwa, Chiba 277-8581, Japan}
\author{Kozo Okazaki}
\affiliation{Institute for Solid State Physics, The University of Tokyo, Kashiwa, Chiba 277-8581, Japan}
\author{Keiichiro Imura}
\affiliation{Department of Physics, Nagoya University, Nagoya, Aichi 464-8602, Japan}
\author{Hiroyuki S. Suzuki}
\affiliation{Institute for Solid State Physics, The University of Tokyo, Kashiwa, Chiba 277-8581, Japan}
\author{Noriaki K. Sato}
\affiliation{Center for General Education, Aichi Institute of Technology, Toyota, Aichi 470-0392, Japan}
\author{Shin-ichi Kimura}
\email{kimura.shin-ichi.fbs@osaka-u.ac.jp}
\affiliation{Graduate School of Frontier Biosciences, Osaka University, Suita, Osaka 565-0871, Japan}
\affiliation{Department of Physics, Graduate School of Science, Osaka University, Toyonaka, Osaka 560-0043, Japan}
\affiliation{Institute for Molecular Science, Okazaki, Aichi 444-8585, Japan}

\date{\today}

\begin{abstract}
The photo-induced band structure variation of a rare-earth-based semiconductor, samarium monosulfide (SmS), was investigated using high-harmonic-generation laser-based time-resolved photoelectron spectroscopy.
A nonlinear photo-induced band shift of the Sm 4\textit{f} multiplets was observed.
The first one is a shift to the high-binding-energy side due to a large surface photovoltage (SPV) effect of approximately 93~meV, comparable to the size of the bulk band gap, with a much longer relaxation time than 0.1~ms.
The second one is an ultrafast band shift to the low binding energy side, which is in the opposite direction to the SPV shift, suggesting an ultrafast valence transition from divalent to trivalent Sm ions due to photo-excitation.
The latter energy shift was approximately 58~meV, which is consistent with the energy gap shift from ambient pressure to the boundary between the black insulator and golden metallic phase with the application of pressure. This suggests that the photo-induced valence transition can reach the phase boundary, but other effects are necessary to realize the golden metallic phase.
\end{abstract}

\maketitle


Materials in the vicinity of critical points, such as strongly correlated electron systems, exhibit characteristic pressure- and magnetic-field-induced phase transitions, such as metal-insulator transitions and magnetically ordered nonmagnetic transitions.

In addition to traditional external perturbations, photo-excitation has recently been developed to create nonequilibrium states where charge-melted metal-insulator transitions \cite{PIMITPerovskite, PIMITVO2} and photo-excited superconducting states \cite{OkazakiPISC, SuzukiPISC} appear.
The photo-induced creation and manipulation of physical properties have recently been studied for future applications in ultrafast electronics\cite{UFQM}. Among the candidate materials, some rare-earth compounds exhibit quantum phase transitions between magnetically ordered states and nonmagnetic heavy fermion (HF) states and valence transitions when applying pressure and magnetic fields \cite{new1, new2, new3}. Such materials are also promising hosts for realizing the photo-excited breakdown of HF states \cite{new4, new5} and the change in mean valences \cite{XAS_EuNiSeGe, UF_EuNiSiGe}.

Samarium monosulfide (SmS) is also located near the valence phase transitions.
At ambient pressure, SmS is a black-colored semiconductor with an indirect energy gap size of 0.1~eV \cite{SmSBG}, and Sm ions are almost divalent (4$f^6$) \cite{SmSproperty}.
Above the critical pressure of approximately 0.65~GPa, this material changes to the golden semimetallic phase together with a valence transition to the trivalent state (4$f^5$) while maintaining the same crystal structure \cite{BGphasetransition1,BGphasetransition2,SmSlatticeconstant1,SmSlatticeconstant2,volumechangelatticekeeping}, namely the black-to-golden phase transition (BGT).
Although the phase transition was discovered more than half a century ago, its origin remains under debate \cite{new6}.
Recent theoretical and experimental studies on the phase transition of SmS have proposed an important role for the generation of electron--hole pairs (excitons) \cite{new7,excitoninstability}.
By the optical excitation of Sm$^{2+}$ 4$f$ electrons, an electronic configuration similar to that in the golden semimetallic phase (4$f^6$ $\rightarrow$ 4$f^5$) with an electron--hole pair can be artificially created, expecting a similar phase transition with the application of pressure.
Therefore, SmS is one of the candidates to investigate the photo-induced nonequilibrium electronic structure, especially whether BGT can be exhibited by photo-excitation.

Time-resolved photoelectron spectroscopy (TrPES) and time- and angle-resolved photoelectron spectroscopy (TrARPES) are powerful experimental tools for investigating the transient electronic structure of the photo-induced state.
The observation of transient band structures by TrARPES has been successful in the discovery of novel inequivalent phenomena, such as Floquet-Bloch states in topological insulators \cite{Flouquet} and an insulator-to-semimetal transition in excitonic insulators \cite{new8}.
However, despite extensive studies on topological materials and low-dimensional systems, observation of the transient band structure in rare-earth compounds has not been carried out until now, except for a few pioneering studies \cite{SPV}.
One of the reasons for the difficulty in observation is that photoelectrons excited by 6-eV photons in ordinary TrARPES have poor 4$f$ photoionization cross sections \cite{crosssection}.

\begin{figure}
\includegraphics[width=80mm]{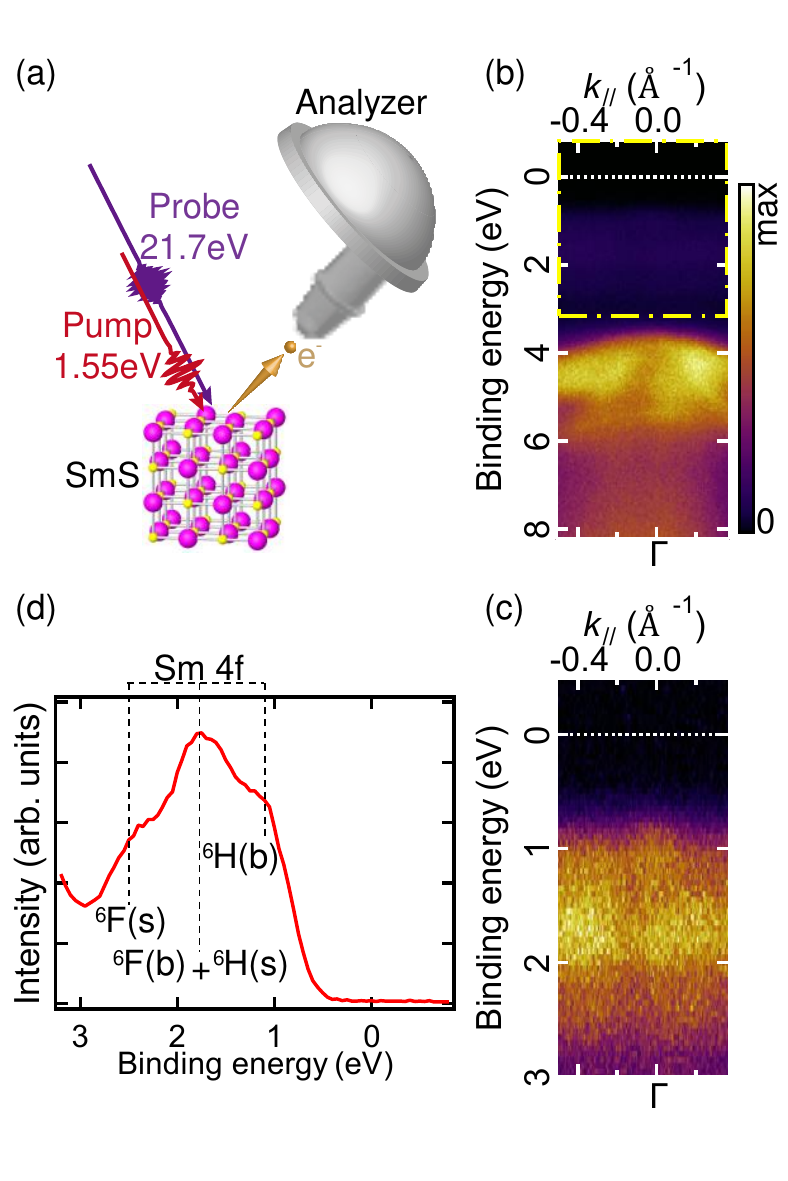}
\caption{\label{figure 1}
(a) Schematic illustration of high-harmonic-generation-laser-based time-resolved photoelectron spectroscopy (TrPES), as applied to SmS.
(b) and (c) band mappings of SmS with angle-resolved photoelectron spectroscopy (ARPES) without the photo-irradiation of a pump laser. (b) Wide-range ARPES image and (c) 4\textit{f} multiplets in the yellow dashed square area of (b).
(d) Angle-integrated photoelectron spectrum of (c). The multiplet structures of the Sm$^{2+}$ 4\textit{f}$^{5}$ final state are visible.
The peak positions of Sm 4\textit{f}$^{5}$ multiplets are indicated by the dashed lines.
}
\end{figure}

In this study, we investigate the photo-induced change in the electronic structure of SmS using TrPES combined with extreme ultraviolet photons ($\sim$22~eV) emitted from a high-harmonic-generation laser. The energy shift of the Sm 4$f$ multiplet structure and modulation of the energy gap in the photo-induced state were successfully visualized.
Two types of characteristic photo-induced energy shifts of the Sm 4\textit{f} multiplet structure were observed.
The first is a large surface photovoltage (SPV) effect of approximately 93 meV on the high-binding-energy side, comparable to the size of the bulk indirect band gap of SmS.
The second is an ultrafast band shift of approximately 58 $\pm$ 4 meV to the low binding energy side, indicating photo-induced band deformation due to photo-excitation from the 4\textit{f} to 5\textit{d} state.
Photo-excitation induces the energy gap narrowing, but the photo-excited phase still has an energy gap, suggesting that the material is not a metal. This suggests that the generation of electron--hole pairs by photo-excitation alone cannot cause BGT, and other effects, such as lattice modulation, are necessary.

\begin{figure}
\includegraphics[width=80mm]{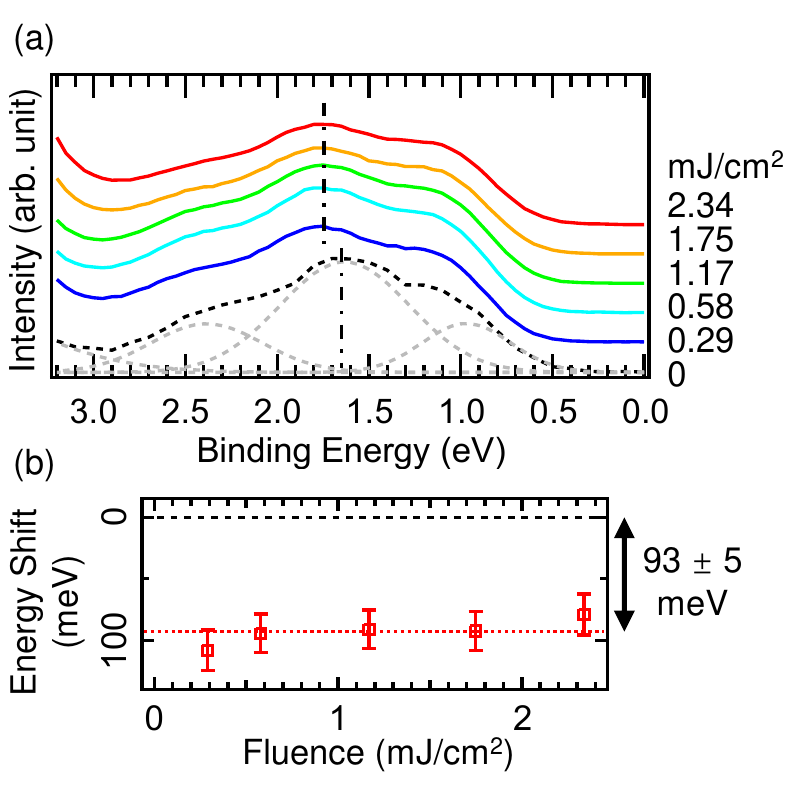}
\caption{\label{figure 2}
(a) Angle-integrated photoelectron spectra of the Sm 4\textit{f} multiplet structure without irradiation of the pump laser (black dashed line) and before the arrival of the pump pulse ($t < 0$) under irradiation of the pump laser with a fluence.
Gray dashed lines indicate the contribution of different parts of the fitting function applied to the photoelectron spectra, and vertical dot-dashed lines indicate the energy positions of the central 4\textit{f} multiplet structure acquired by fitting.
(b) Pump fluence dependency of the energy shift of the 4\textit{f} multiplet structure happens at t $<$ 0. Fitting results obtained using a horizontal line are shown using a red dashed line.
}
\end{figure}

\begin{figure*}
\includegraphics[width=150mm]{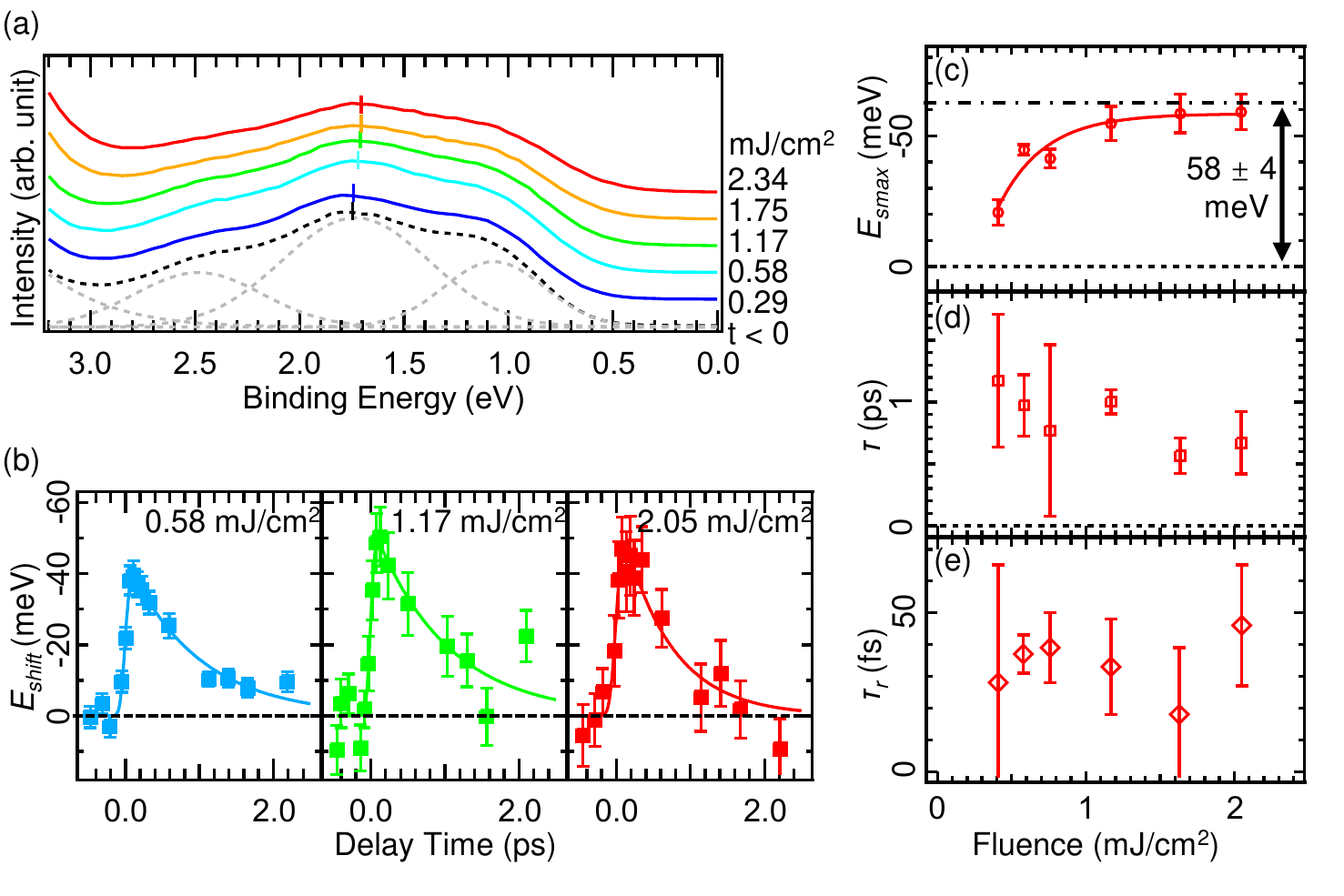}
\caption{\label{figure 3}
 Angle-integrated photoelectron spectra of the multiplet structure of SmS 4\textit{f} final state before the arrival of the pump pulse ($t<0$) (black dashed line) and just after the arrival of the pump pulse (less than 0.1 ps) under irradiation of the pump laser with a fluence. Similar to the indications in Fig. 2(a), gray dashed lines represent three Sm $4f$ multiplet peaks, and short vertical lines indicate the approximate positions of the central peaks.
 (b) Time structure of the energy shift of the three 4\textit{f} multiplet peaks at various pump fluences. Marks show the energy shift evaluated by the fitting of photoelectron spectra, while the solid lines show the results of a single-exponential decay function convoluted with a Gaussian function fitting.
 (c) Pump fluence dependence of the maximum energy shift of the 4\textit{f} multiplet structure at $t>0$. The red solid line shows the fitting result with a single-exponential decay function, while the horizontal dot-dashed line shows the saturation value of the energy shift.
 (d) Relaxation time of the fitting function in (b) as a function of pump fluence.
 (e) Energy shift time after photo-irradiation as a function of pump fluence.
}
\end{figure*}

High-quality single-crystalline SmS was synthesized using the vertical Bridgman method in a high-frequency induction furnace.
A clean SmS surface was obtained by cleaving \textit{in situ} in the measurement chamber under a background pressure of approximately 10$^{-9}$ Pa.
Figure 1(a) illustrates the experimental system of TrARPES.
The pump pulse is 1.55 eV light generated directly from a commercial Ti:Sapphire regenerative amplifier system (Spectra-Physics, Solstice Ace), whereas the probe pulse is 21.7 eV light produced by the seventh harmonic generation of the second harmonic, which is also used from the identical laser of the pump pulse \cite{SuzukiTRPES}.
The repetition rate and pulse width of the pump pulse were 10 kHz and 35 fs, respectively, whereas the temporal and energy resolutions were estimated to be about 70 fs and 250 meV, respectively.
Photoelectrons are detected by using a hemispherical electron analyzer (Scienta, R4000).
All experiments were conducted at room temperature.
Here, we mainly focused on the energy shift of weakly dispersive Sm 4\textit{f} multiplet peaks; therefore, we integrated the TrARPES image along the momentum direction to obtain angle-integrated photoelectron (AIPE) spectra with a high S/N ratio.

Figure 1(b) shows the wide-range ARPES intensity plot of SmS without photo-irradiation by the pump laser.
Dispersive bands appeared at $\sim$4~eV, and flat bands appeared at $\sim$1.7~eV, originating from the S 3\textit{p} bands and Sm 4\textit{f} multiplets, respectively.
Although the Sm 5$d$ bands are located above the Fermi level at the X point in the bulk Brillouin zone, no such conduction bands are observed, even immediately after photo-excitation.
The possible reasons for the absence of such conduction bands might be the ultrafast relaxation of the excited carrier of less than 70~fs, the time resolution, and/or the difference in the out-of-plane wave number $k_z$.
By setting the inner potential to 14.1~eV \cite{innerP}, the value of $k_z$ was evaluated to be 2.9 $\text{\AA}^{-1}$ for a 21.7~eV probe photon, which is located near the center between the $\Gamma$ point and X point of the bulk Brillouin zone of SmS, while the bottom of the Sm 5$d$ band is located at the X point.
The $k_z$ broadening was estimated to be 0.1~\AA~($\sim$10~$\%$ of the distance from the $\Gamma$ point to K point); thus, an average wavenumber over a wide $k_z$ area was observed, which could result in the indistinctness of the strongly dispersive Sm 5$d$ band.

Figures 1(c) and 1(d) show a magnified ARPES image and the corresponding AIPE spectrum.
The three peaks at approximately 1, 1.7, and 2.5~eV indicated by dashed lines in Fig. 1(d) are the Sm 4\textit{f} multiplets of bulk-$^6H$ ($^6H$(b)), the overlap of bulk-$^6F$ ($^6F$(b)) and surface-$^6H$ ($^6H$(s)), and surface-$^6F$ ($^6F$(s)), respectively.
The shape and position of these peaks are consistent with the previous ARPES data probing by He-\uppercase\expandafter{\romannumeral1} (21.2 eV) photons \cite{SmS-ARPES}, indicating that the valence band structure of SmS can be observed by extreme ultraviolet pulse laser excitation.

\begin{figure*}
\includegraphics[width=150mm]{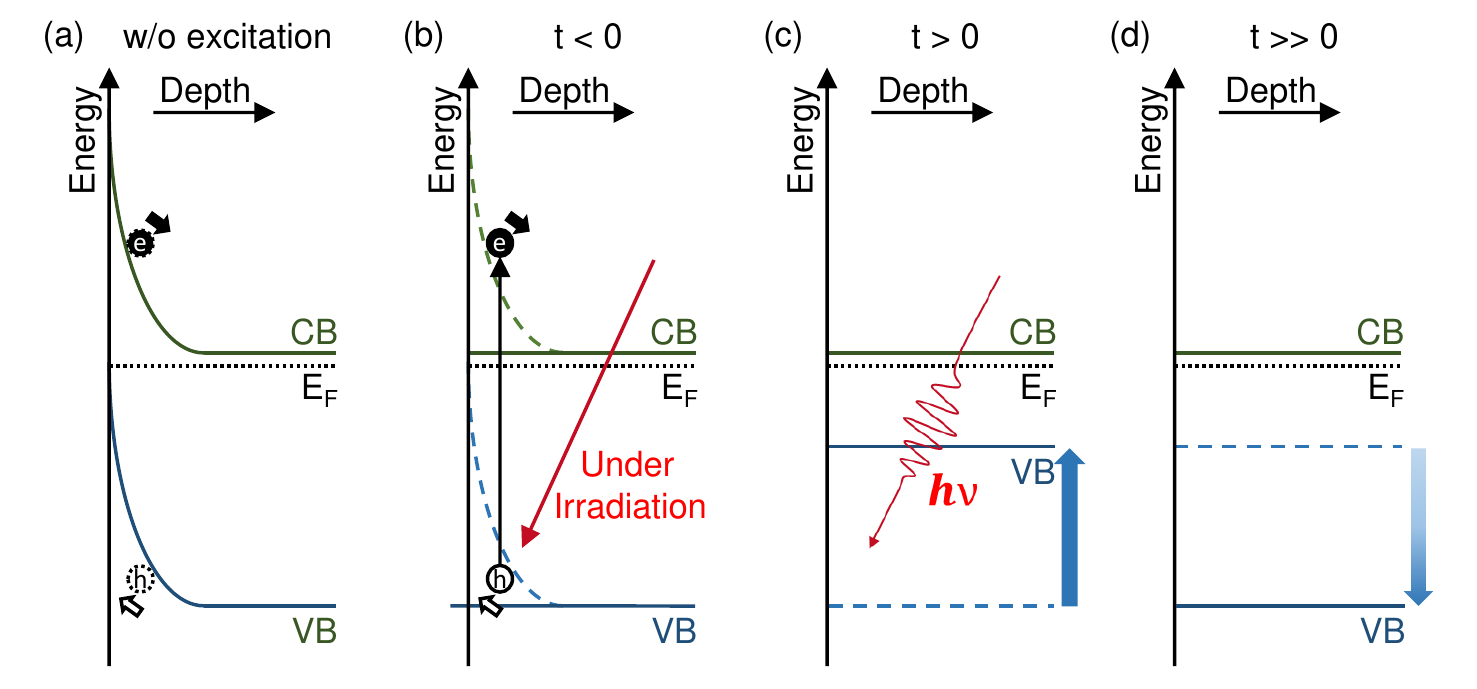}
\caption{\label{figure 4}
Schematic illustration of the band structure near the surface of the SmS sample with photo-irradiation.
(a) Without pumping.
(b) Under photo-irradiation by pump pulses and before the arrival of the pump pulse.
(c) Within several tens of femtoseconds after the arrival of a pump pulse.
(d) Several picoseconds after the arrival of a pump pulse.
}
\end{figure*}

Figure 2(a) shows the AIPE spectra of the multiplet structure of Sm$^{2+}$ 4\textit{f}$^{5}$ final state before the arrival of the pump pulse ($t < 0$).
Even at $t < 0$, the valence band spectrum is shifted by photo-irradiation.
To determine the exact energy positions of the multiplet peaks, the AIPE spectra were fitted using the following function:

\begin{displaymath}
I(x)=\sum^{3}_{i=1}I_{i}\text{exp}(-\frac{(x-x_{i})^{2}}{w_{i}^{2}})+I_{b}\text{exp}(-\frac{(x-x_{b})^{2}}{w_{b}^{2}})+BG
\end{displaymath}
The fitting function consists of four Gaussian functions and a constant background. The first three Gaussian functions with subscript $i$ are for the multiplet structure of the Sm$^{2+}$ 4\textit{f}$^{5}$ final state, as shown in Fig. 1(d), and the other with subscript $b$ is for a structure located at a higher binding energy. \textit{I}, \textit{x}, and \textit{w} are the fitting parameters of intensity, central binding energy, and full width at half maximum of the corresponding Gaussian peak, respectively.

To obtain the overall chemical shift due to photo-irradiation, the widths of the Gaussian peaks representing the 4\textit{f} multiplet structure ($w_{i}$), as well as the energy differences among the three multiplet peaks, were kept unchanged during the fitting process, while other parameters were allowed to vary.
The fitting result of the spectrum without pump laser irradiation is shown by gray dashed lines in Fig. 2(a), and the energy positions of the central Gaussian peak are marked by vertical dot-dashed lines.
After fitting to all other spectra, the size of the peak shift can be evaluated as a constant value of $95\pm5$~meV for all photo-irradiation fluences, as shown in Fig. ~2(b). The peak shift energy is saturated even at the lowest fluence of 0.29 J/cm$^2$.
This energy shift could be attributed to the long-living surface photovoltage (SPV) effect. Further details on this energy shift will be provided later.

Figure 3(a) shows the AIPE spectra of the multiplet structure of the Sm$^{2+}$ 4\textit{f}$^5$ final state, obtained immediately before the arrival of the pump pulse (black dashed line spectrum is for $t < 0$), and the spectra obtained immediately after the arrival (colored solid lines labeled as different fluences) of the pump pulse.
The spectra were fitted to the same function as in Fig. 2(a), and the energy position of the central Gaussian peak of the 4\textit{f} multiplet structure is marked by short vertical lines in each spectrum.
After photo-irradiation, the spectrum shifts toward low binding energies but in the opposite direction to the shift at $t<0$, that is, a nonlinear shift in time.
To investigate the time development of the energy shift after photoirradiation, the same fitting procedure as shown in Fig. 2(a) was performed on the obtained data, and the peak shift was plotted as a function of delay time, as shown in Fig. 3(b).
The AIPE spectrum shifted towards the Fermi level within 0.1 ps and then gradually shifted back to the position before pumping in the following several picoseconds.
To obtain a quantitative understanding, we fitted the time-dependent energy shift to a single-exponential decay function convoluted with a Gaussian function, as shown in Fig. 3(b). The fitting parameters, that is, the maximum energy shift ($E_{smax}$), recovery time $\tau$, and rise time \textit{$\tau_r$} at different pump fluences, are plotted as functions of the pump fluence in Figs. 3(c)--(e).

Figure 3(c) shows the pump fluence dependency of $E_{smax}$. $E_{smax}$ initially increases as the fluence increases up to a pump fluence of approximately 1.2 mJ/cm$^2$ and then saturates at a certain value.
The saturation value of $E_{smax}$ was evaluated as $58\pm4$~meV by assuming a single-exponential function.
The recovery time $\tau$ and rising time \textit{$\tau_r$} are shown in Figs. 3(d) and 3(e), respectively.
Both parameters remained constant at each fluence; $\tau$ was as long as 1~ps, and $\tau_r$ was several tens of femtoseconds, which is comparable to the time resolution of the instrument.
This ultrafast band shift and recovery are believed to correspond to the valence transition of Sm ions, as discussed later.

We now discuss the band shift induced by photo-excitation.
SmS is expected to be an $n$-type semiconductor; thus, a potential point from the bulk to the surface is created by the electrons trapped by the surface state, and the bands near the surface bend upwards without photo-irradiation, as shown in Fig. 4(a).
Under photo-irradiation, excited electrons and holes move to the bulk and surface sides, respectively, driven by the electric field near the surface, which is known as the SPV effect.
Consequently, the surface electric field is canceled when the irradiation power is sufficient.
The observed energy shift to the high-binding-energy side at $t < 0$ can be explained by the cancelation of surface band bending by the SPV effect.
It should be noted that the energy shift before a pump pulse arrives suggests an extremely long lifetime of the SPV effect of at least 0.1 ms, according to the repetition rate of 10 kHz.
A similar long-living SPV shift has been reported for SmB$_6$ \cite{SPV}, but to the best of our knowledge, it does not appear in conventional semiconductors \cite{new10}.
The lifetime of the SPV effect may originate from the nondispersive Sm 4\textit{f} band structure, which is a common feature of SmS and SmB$_6$.

By the pump pulse, localized 4\textit{f} electrons of Sm ions were excited to the widely spread 5\textit{d} conduction band. Then, the ionic valence of Sm changed from a divalent to nearly trivalent state, and the band structure exhibited a trend of changing from a divalent to trivalent state, as in Ref. \cite{new9}, that is, the 5\textit{d} conduction band is expected to shift downward, while the 4\textit{f} valence band shifts upward, thus reducing the band gap, as shown in Fig. 4(c); finally, the band shift returned to the same value as that at the SPV effect (Fig. 4(d)).
Because the peak shift energy due to the SPV effect at $t < 0$ ($93\pm5$~meV) is comparable to the bandgap size of SmS ($93\pm4$~meV) \cite{excitoninstability}, the bottom of the 5\textit{d} conduction band is considered to be located just above the Fermi level at the X point.
Furthermore, no evident change in the sample color was observed during the time-resolved measurement and the 5\textit{d} occupied state was not observed in the ARPES mapping after photo-irradiation. Therefore, the bottom of the 5\textit{d} band was pinned at the Fermi level even after photo-irradiation, meaning that SmS continued to be a semiconductor.
However, the photo-induced shift of the 4\textit{f} valence band toward the Fermi level shown in Fig. 3 suggests the narrowing of the band gap.
Considering the indirect band gap size of bulk SmS and the saturation value of $E_{smax}$ of $58\pm4$~meV, the saturation value of the energy gap size ($\sim$35~meV) under photo-irradiation is similar to that ($\sim$45~meV) observed just before the BGT critical pressure of 0.65~GPa is attained \cite{excitoninstability}.
In BGT, the energy gap suddenly collapsed across the critical pressure.
However, the gap size of the photo-excitation state saturated to the value at the critical pressure, suggesting that the band shift only occurred in the black phase and did not cross the phase boundary of the BGT.
One of the reasons for this is that the photo-irradiation time ($\sim$ps) is too short for the lattice to shrink from the black phase to the golden phase ($>$ ns).

In summary, we performed TrPES on SmS and experimentally specified two individual energy shifts of the multiplet structure (4\textit{f}$^6$ $\rightarrow$ 4\textit{f}$^5$) of Sm$^{2+}$ induced by photo-excitation. At $t<0$, we observed an energy shift to the high-binding-energy side, which was attributed to the SPV effect.
At $t>0$, the 4\textit{f} multiplet structure shifted to the low binding energy side within several tens of femtoseconds and recovered completely in a few picoseconds. This suggests that creating excitons by photo-excitation exhibits the valence transition of SmS in a short time, but the phase does not change to the golden phase, which appears at a high pressure.
This further suggests that the creation of excitons is the origin of the narrowing of the energy gap, which supports the theory that excitons play an important role in BGT.
However, the short period of appearance of excitons did not cause BGT.
Other effects, including lattice shrinking, are considered to be necessary for BGT.

We would like to thank Professor Takahiro Ito for helpful discussions.
The TrARPES experiment in this work was conducted by joint research at the ISSP, University of Tokyo.
This study was supported by JSPS KAKENHI (Grant Nos. 20H04453 and 22K14605) and the Iketani Science and Technology Foundation.


\bibliography{reference.bib}

\newpage
\title{Supplementary Information: }


\end{document}